\documentclass[lettersize,journal]{IEEEtran}
\usepackage{amsmath,amsfonts}
\usepackage{algorithmic}
\usepackage{algorithm}
\usepackage{array}
\usepackage[caption=false,font=normalsize,labelfont=sf,textfont=sf]{subfig}
\usepackage{textcomp}
\usepackage{stfloats}
\usepackage{url}
\usepackage{verbatim}
\usepackage{graphicx}
\usepackage{cite}

\usepackage{graphicx} 
\usepackage{multirow}
\usepackage{amsmath}
\usepackage{adjustbox}
\usepackage{amsthm}

\usepackage{booktabs}
\usepackage{makecell}
\usepackage{amssymb}
\begin{document}

\title{PRISM: Refracting the Entangled User Behavior Space for E-Commerce Search}

\author{Haoqian Zhang, Ziyuan Yang, \IEEEmembership{Member, IEEE}, and Yi Zhang, \IEEEmembership{Senior Member, IEEE}
\thanks{Haoqian Zhang, and Yi Zhang are with the School of Cyber Science and Engineering, Sichuan University, Chengdu 610065, China (e-mail: 2023141490116@stu.scu.edu.cn; yzhang@scu.edu.cn).}
\thanks{Ziyuan Yang is with the Lee Kong Chian School of Medicine, Nanyang Technological University, Singapore (e-mail: cziyuanyang@gmail.com).}
}



\maketitle

\begin{abstract}
E-commerce search systems rely on modeling user behavior to estimate item relevance and user preference, which are typically assumed to be stable and independently learnable signals. However, in practice, user interactions are jointly shaped by exposure mechanisms, feedback loops, and semantic matching, leading to entangled and dynamically drifting behavioral signals. As a result, both preference estimation and relevance modeling suffer from confounding effects and semantic misalignment, which limits the robustness of downstream ranking models. To address this issue, we propose PRISM, a Preference–Relevance Interaction Semantic Modeling framework for e-commerce search behavior prediction. PRISM explicitly models the interaction between user preference and item relevance rather than treating them as independent components. Specifically, it introduces a preference rectification module to iteratively refine user preference under relevance-aware constraints, improving robustness against behavioral confounding. To ensure semantic consistency, we further incorporate a large language model (LLM)-driven semantic anchoring mechanism that leverages positive and negative prototypes to calibrate relevance representations. Finally, a preference-conditioned evidence routing module adaptively aggregates multi-source behavioral signals, enabling context-aware and preference-aligned relevance estimation.
Extensive experiments on two public e-commerce benchmarks demonstrate that PRISM consistently outperforms strong baselines, validating the effectiveness of explicitly modeling preference–relevance interaction for robust and semantically grounded search behavior modeling.
\footnote{Our code will be made publicly available after the review process.}
\end{abstract}


\section{Introduction}
E-commerce has reshaped the retail industry by allowing consumers to browse and purchase a wide variety of products and services at any time through online platforms~\cite{Tsagkias2020,Sun2023KuaiSAR}. In this context, search plays a central role by retrieving and ranking relevant items in response to user queries, thereby directly influencing user engagement~\cite{Xu2024GRACE,Li2019SemanticPairwise}. Consequently, a central goal of e-commerce search modeling is to accurately understand user behavior, particularly why users interact with some displayed certain items over others ~\cite{Wang2025DRP,Liu2023JDsearch,Shi2024UniSAR}. 

Recent studies have increasingly focused on jointly modeling item relevance and user preference to understand user behavior~\cite{Ai2019ZeroAttention,Bi2021ReviewTransformer}. Item relevance captures the semantic alignment between a query and an item, user preference reflects user-specific interest that go beyond query-dependent matching. For example, PRECTR integrates item relevance matching and click-through rate (CTR) prediction into a unified framework~\cite{Chen2025PRECTR}. PRINT further models the personalized incentive of query-ad semantic relevance on click probability~\cite{Hong2024PRINT}. These works have shown that although item relevance and user preference are important in understanding user behavior, effectively modeling their interplay remains challenging~\cite{Luo2024CausalSurvey}.

These methods inherently assume that the relationship between item relevance and user preference is static and well-defined, which contradicts practical settings. User behavior does not correspond to stable semantic representations, but rather to indirect observations jointly shaped by the exposure mechanism, interaction feedback, and semantic similarity. Consequently, user preference does not necessarily reflect a user’s true and stable interests, but instead represents a mixed signal influenced by multiple confounding factors~\cite{Luo2024CausalSurvey,Lin2024ReCRec}. Similarly, item relevance is not a purely semantic matching signal, as it is also affected by behavioral feedback, leading to semantic drift~\cite{Niu2024CLTRRobustness,Coppolillo2025AlgorithmicDrift}. Such semantic drift means that relevance estimation gradually deviates from the true query--item semantic matching. Hence, a key challenge lies in disentangling and dynamically modeling item relevance and user preference under such entangled and confounded observations.

Motivated by the above observations, we argue that effective user behavior modeling should move beyond treating item relevance and user preference as static or independently learnable signals~\cite{Luo2024CausalSurvey,Liu2024CausalTrustworthySurvey}. Instead, it requires capturing their dynamic interplay and contextual variability, where observed behaviors reflect a mixture of semantic intent and interaction-induced behaviors. This calls for modeling behavioral signals in a way that can maintain semantic fidelity in relevance estimation while accounting for the inherent uncertainty and heterogeneity in user preference. First, the modeling process should be robust to confounding effects introduced by the interaction, which may obscure the true user preference signal~\cite{Lin2024ReCRec,Liang2024DeconfoundingPreference}. Second, item relevance estimation should remain semantically grounded in the query–item matching relationship under dynamic behavioral feedback, rather than drifting toward spurious patterns reinforced by recursive interactions~\cite{Niu2024CLTRRobustness,Coppolillo2025AlgorithmicDrift}. Third, it should be capable of adaptively utilizing heterogeneous evidence in a way that is consistent with the current behavioral context, rather than treating all signals uniformly~\cite{Bi2021ReviewTransformer,Liu2022KDCRM,Liu2025MRSurvey}.

Specifically, in this paper, to address the above issues, we propose a \textbf{P}reference–\textbf{R}elevance \textbf{I}nteraction \textbf{S}emantic \textbf{M}odeling~(\textbf{PRISM}) method for behavior prediction in e-commerce search. PRISM consists of three components. First, we introduce a preference rectification module that formulates preference refinement by explicitly modeling its interaction with relevance signals, enabling controllable and instance-adaptive preference purification. Then, we develop a large language model~(LLM)-driven semantic anchoring module to improve item relevance estimation through semantic calibration. Specifically, positive and negative prototypes derived from LLM-generated priors are introduced as external semantic anchors. These anchors provide structured guidance for calibrating the semantic polarity of relevance representations, thereby improving discriminative alignment. Finally, we design a preference-conditioned evidence routing mechanism that selectively aggregates multi-source evidence, which is derived from the query, user, target item, and their pairwise interactions, under the edited preference state and uses the routed evidence to refine relevance estimation. In this way, our method can effectively transform noisy and entangled user behavior signals into a structured, preference-calibrated semantic space, enabling more robust and semantically aligned relevance modeling in e-commerce search. 

Our contributions are summarized as follows:
\begin{itemize}
\item We study the intrinsic limitation of entangled user behavior modeling and propose a novel Preference–Relevance Interaction Framework~(PRISM) for behavior prediction in e-commerce search.
\item We introduce a LLM-driven semantic constraint module to improve item relevance estimation through semantic calibration.
\item We propose a preference-conditioned evidence routing mechanism to enhance item relevance estimation through adaptive evidence selection.
\item We conduct extensive experiments on two public benchmarks, demonstrating the effectiveness of the proposed framework over strong baselines.
\end{itemize}

\section{Related Work}
\label{sec:related}

\subsection{Personalized Product Search and User Behavior Modeling}

Personalized product search aims to rank items by jointly considering the current query and the user's historical preferences. Early studies mainly focused on learning user-aware matching functions for query~\cite{Li2019SemanticPairwise,Yao2021CLK,Liu2022KDCRM}. For example, \textit{Ai et al.}~\cite{Ai2017HEM} proposed a hierarchical embedding framework for personalized product search, while \textit{Guo et al.}~\cite{Guo2019ALSTP} modeled long-term and short-term preferences to capture users' evolving interests. \textit{Ai et al.}~\cite{Ai2019ZeroAttention} further investigated when personalization should be activated for a query, and \textit{Bi et al.}~\cite{Bi2020TEM}, \textit{Bi et al.}~\cite{Bi2021ReviewTransformer} introduced Transformer-based architectures to better encode user reviews and fine-grained behavioral evidence. These studies have demonstrated the importance of incorporating personalized information into search ranking, but they typically focus on improving user-aware relevance estimation rather than explicitly characterizing the distinct roles of relevance and preference.

Related efforts in click-through rate prediction and user behavior modeling have also emphasized the importance of user interest representations~\cite{Zhou2018DIN,Zhou2019DIEN,Chen2019BST}. More broadly, neural CTR models such as \textit{Guo et al.}~\cite{Guo2017DeepFM} and \textit{Wang et al.}~\cite{Wang2017DCN} have shown that feature interaction modeling is critical for response prediction. However, these methods generally optimize the final behavioral label directly and often treat the observed interaction as a unified supervision signal, without explicitly disentangling query-dependent relevance from user-specific preference in search impressions.

\subsection{Joint Modeling of Relevance and Preference in Search}

Recent research has increasingly recognized that user behavior in search cannot be adequately explained by relevance or preference alone. \textit{Carmel et al.}~\cite{Carmel2020VoiceRelevance} showed that customer satisfaction in e-commerce search is not always aligned with conventional relevance judgments, suggesting that behavioral outcomes may reflect additional personalized factors beyond semantic matching. Along this line, \textit{Hong et al.}~\cite{Hong2024PRINT} modeled personalized relevance incentive in sponsored search CTR prediction, and \textit{Chen et al.}~\cite{Chen2025PRECTR} proposed a unified framework that integrates personalized search relevance matching with CTR prediction. More recently, \textit{Wang et al.}~\cite{Wang2025DRP} pointed out that user behavior in e-commerce search is jointly shaped by relevance effects and preference effects, and proposed reconstructing the behavior modeling space to mitigate their entanglement.

These studies provide strong evidence that jointly modeling relevance and preference is crucial for search behavior prediction. Nevertheless, existing methods mostly formulate the two factors through static fusion or unified supervision, and less attention has been paid to how preference should be purified under entangled feedback, how relevance should remain semantically grounded under behavioral drift, and how the interaction between the two should be adaptively modeled at the instance level.

\subsection{Debiasing and Deconfounding from Logged Interactions}

Learning from logged interactions is inherently challenging because behavioral signals are shaped not only by user interests and item relevance, but also by exposure bias, position bias, popularity effects, and other confounding factors. In the ranking literature, \textit{Joachims et al.}~\cite{Joachims2017ULTR} established the counterfactual perspective for learning from biased implicit feedback, and subsequent work has continued to investigate the robustness and limitations of counterfactual learning-to-rank methods~\cite{Gupta2023ULTR,Niu2024CLTRRobustness}. In recommendation, causal and deconfounded learning has become an important direction for handling biased observations~\cite{Wang2020CausalRecSys,Luo2024CausalSurvey,Liu2024CausalTrustworthySurvey}. For example, \textit{Lin et al.}~\cite{Lin2024ReCRec} reasoned about the causes of implicit feedback for debiased recommendation, while \textit{Liang et al.}~\cite{Liang2024DeconfoundingPreference} explicitly studied how to deconfound user preference using both implicit and explicit feedback.

These findings suggest that logged behavior should not be interpreted as a direct reflection of true preference or true relevance. Instead, it is a mixed signal shaped by multiple mechanisms. Moreover, repeated interaction between models and users may gradually distort the semantics of learned signals, leading to preference shift or semantic drift~\cite{Ge2020EchoChambers,Coppolillo2025AlgorithmicDrift}. While existing debiasing and deconfounding methods mainly focus on statistical correction or causal identification, they are less concerned with preserving semantically meaningful relevance estimation while simultaneously refining preference representations in search scenarios.

\subsection{Semantic Enhancement with Pre-trained and Large Language Models}

Another related line of work seeks to improve behavior prediction by incorporating richer semantic knowledge from pre-trained language models or large language models. In sponsored search, \textit{Wang et al.}~\cite{Wang2022NLPCTR} introduced supplementary NLP features to enhance CTR prediction beyond conventional sparse features. More recent studies have explored tighter integration between language models and CTR architectures. For instance, \textit{Lin et al.}~\cite{Lin2024ClickPrompt} used CTR models as prompt generators for adapting language models to CTR prediction, and \textit{Geng et al.}~\cite{Geng2024BAHE} investigated how to enhance CTR modeling with large language models for long textual user behaviors. In recommendation, \textit{Xi et al.}~\cite{Xi2024KAR} proposed augmenting recommendation models with external knowledge distilled from large language models, and recent surveys have summarized the broader trend of LLM-enhanced recommender systems~\cite{Liu2024LLMRSurvey}.

Although these approaches demonstrate the value of external semantic priors, most of them use language models as feature enhancers, knowledge providers, or auxiliary encoders. Comparatively less effort has been devoted to using external semantic priors to explicitly calibrate the semantic polarity of relevance representations in search. This leaves open the question of how to introduce lightweight yet effective semantic grounding into relevance modeling while maintaining compatibility with preference-aware behavior prediction.

Overall, prior work has advanced personalized product search, joint relevance--preference modeling, debiasing from logged feedback, and semantic enhancement for behavior prediction. However, there remains limited research on a unified search behavior modeling framework that simultaneously rectifies entangled preference signals, semantically calibrates relevance representations, and adaptively models their interaction under logged search impressions.

\section{Problem Statement}
\label{sec:problem}

Let $\mathcal{D}=\{(x_i,y_i)\}_{i=1}^{N}$ denote the logged search dataset, where $N$ is the number of logged impressions. For the $i$-th impression, $x_i$ can be represented as a triplet consisting of the user context $u_i$, query $q_i$, and exposed item $v_i$, which can be formulated as $x_i=(u_i,q_i,v_i)$. The label $y_i\in\{0,1\}$ denotes the observed behavior label, where $y_i=0$ means no interaction with the search results and $y_i=1$ means that the user interacted with at least one search result. 
The goal of CTR prediction is to estimate a behavior prediction score $\hat{y}_i$ that closely approximates the true label $y_i$.

In e-commerce search, the user behavior is jointly determined by two key factors: query-item relevance and user preference. Query-item relevance reflects whether the exposed item matches the search intent implied by $q_i$. The user preference characterizes whether the item is attractive to the user under the current search context. Accordingly, joint modeling framework models these two factors using a preference model $\mathrm{PM}$ and a relevance model $\mathrm{RM}$.

Based on the above formulation, the behavior prediction task in search can be viewed as jointly learning a preference model $\mathrm{PM}$, a relevance model $\mathrm{RM}$, and an interaction function $g$. The interaction function combines personalized preference and query-item relevance for behavior estimation. Then, the overall learning objective is to minimize the empirical loss over all impressions:
\begin{equation}
\label{eq:problem_objective}
\min_{\mathrm{PM},\mathrm{RM},g}
\frac{1}{N}\sum_{i=1}^{N}
\ell\!\left(
y_i,\,
g\big(\mathrm{PM}(u_i,q_i,v_i),\mathrm{RM}(q_i,v_i)\big)
\right),
\end{equation}
where $\ell(\cdot,\cdot)$ is the loss function for behavior prediction.

More specifically, the relevance score is defined as
$\hat{r}_i=\mathrm{RM}(q_i,v_i)$, which captures the matching degree between the query and the exposed item. The preference score is defined as
$\hat{p}_i=\mathrm{PM}(u_i,q_i,v_i)$, where the query is also incorporated into the preference model to capture the user's intent in the current search session. Then, the final behavior prediction is written as:
\begin{equation}
\label{eq:problem_factorization}
\hat{y}_i
=
g(\hat{p}_i,\hat{r}_i)
=
g\big(\mathrm{PM}(u_i,q_i,v_i),\mathrm{RM}(q_i,v_i)\big).
\end{equation}


\begin{figure*}[!t]
    \centering
    \includegraphics[width=\textwidth]{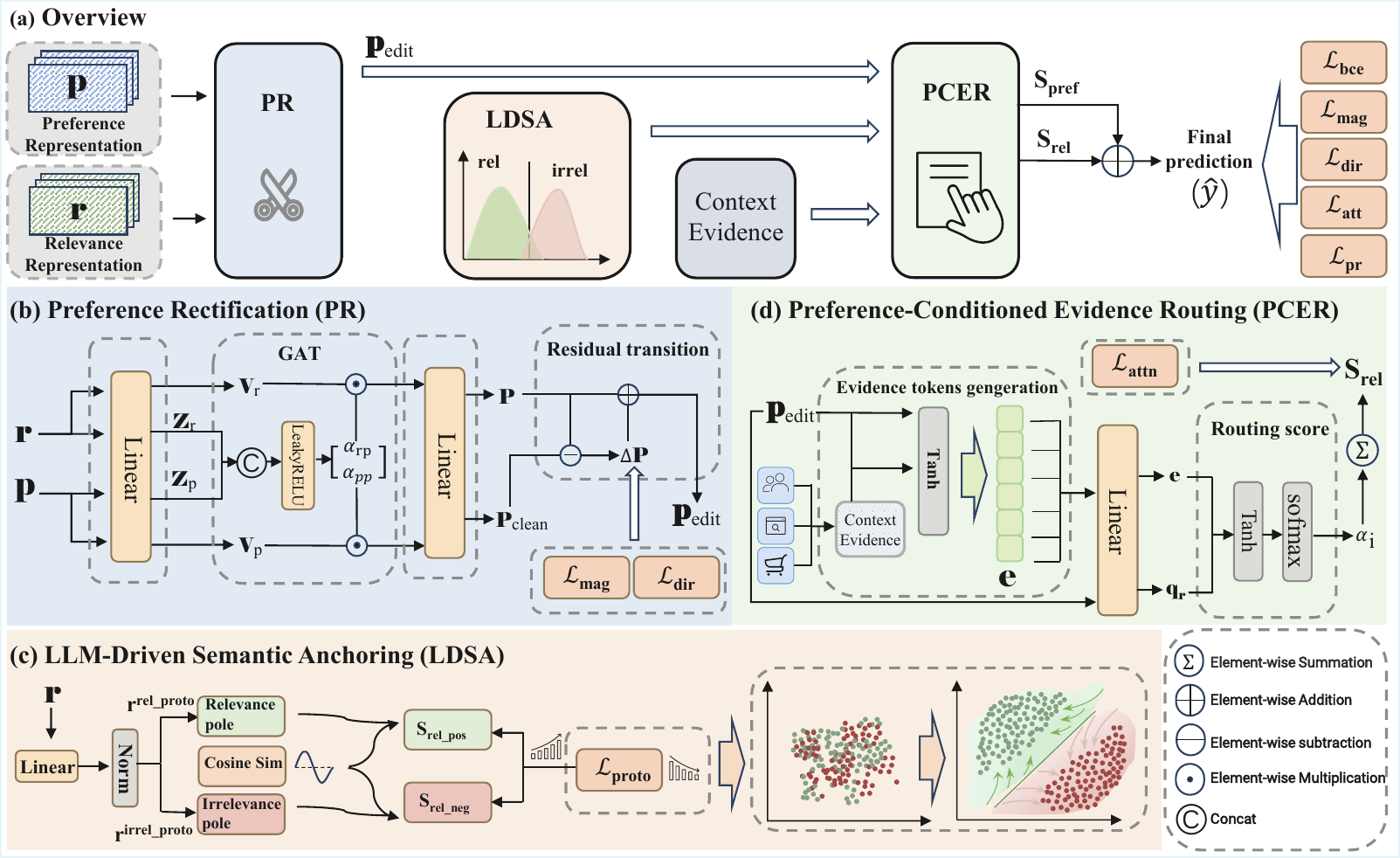}
    \caption{The overview of our proposed method.}
    \label{fig:framework}
\end{figure*}

\section{Method}
\label{sec:method}

\subsection{Overview}
\label{subsec:overview}

In this paper, we propose a \textbf{P}reference--\textbf{R}elevance \textbf{I}nteraction \textbf{S}emantic \textbf{M}odeling (\textbf{PRISM}) framework for search behavior prediction. The overall framework of PRISM is illustrated in Figure~\ref{fig:framework}. PRISM consists of three components: \emph{Preference Rectification}, \emph{LLM-Driven Semantic Anchoring}, and \emph{Preference-Conditioned Evidence Routing}. Preference Rectification aims to obtain a robust preference representation by suppressing relevance-induced interference while preserving the original preference semantics. LLM-Driven Semantic Anchoring regularizes the relevance representation with fixed semantic prototypes derived from the priors. Preference-Conditioned Evidence Routing further refines relevance modeling by selectively aggregating task-adaptive evidence conditioned on the rectified preference state. Finally, the behavior prediction is obtained by an interaction function that combines the preference score and the refined relevance score.

\subsection{Preference Rectification}
\label{subsec:prism}

Let $\mathbf{p}\in\mathbb{R}^{d}$ and $\mathbf{r}\in\mathbb{R}^{d}$ denote the preference representation and relevance representation, respectively. $d$ is the hidden dimension. Then, $\mathbf{p}$ is produced by the preference model, while $\mathbf{r}$ is produced by the relevance model. To model their interaction, we project them into a shared attention space: $\mathbf{z}_{p}=\mathbf{W}_{s}\mathbf{p}$, $\mathbf{z}_{r}=\mathbf{W}_{s}\mathbf{r}$, $\mathbf{v}_{p}=\mathbf{W}_{v,p}\mathbf{p}$, and $\mathbf{v}_{r}=\mathbf{W}_{v,r}\mathbf{r}$. Here, $\mathbf{W}_{s}$ is the shared projection matrix for attention scoring, $\mathbf{W}_{v,p}$ and $\mathbf{W}_{v,r}$ are value projection matrices, $\mathbf{z}_{p}$ and $\mathbf{z}_{r}$ are projected representations for attention computation, and $\mathbf{v}_{p}$ and $\mathbf{v}_{r}$ are value representations.

We then compute the attention scores of the relevance-to-preference path and the self-preserving preference path:
\begin{equation}
\label{eq:prism_scores}
\begin{aligned}
e_{rp} &= \mathrm{LeakyReLU}
\left(\mathbf{a}^{\top}[\mathbf{z}_{r};\mathbf{z}_{p}]\right),\\
e_{pp} &= \mathrm{LeakyReLU}
\left(\mathbf{a}^{\top}[\mathbf{z}_{p};\mathbf{z}_{p}]\right),
\end{aligned}
\end{equation}
where $e_{rp}$ and $e_{pp}$ denote the relevance-to-preference attention score and the self-preserving attention score, respectively. $\mathbf{a}$ is a learnable attention vector, $[\cdot;\cdot]$ denotes the concatenation operation. $\mathrm{LeakyReLU}(\cdot)$ is the activation function. The corresponding normalized coefficients are obtained by $[\alpha_{rp},\alpha_{pp}]=\mathrm{softmax}([e_{rp},e_{pp}])$, where $\alpha_{rp}$ and $\alpha_{pp}$ are the attention weights of the two paths.

The two normalized coefficients can be interpreted as an adaptive gate between relevance-induced influence and preference-preserving information. Since the goal of Preference Rectification is to reduce the interference from relevance while preserving the intrinsic preference semantics, we use the self-preserving coefficient $\alpha_{pp}$ to modulate the preference value representation. The rectified preference message is then formulated as $\mathbf{m}_{p}^{\mathrm{clean}}=\alpha_{pp}\mathbf{v}_{p}$. Based on this message, we obtain a rectified preference candidate by $\mathbf{p}_{\mathrm{clean}}=\mathbf{W}_{o}\sigma(\mathbf{m}_{p}^{\mathrm{clean}})$, where $\mathbf{W}_{o}$ is an output projection matrix and $\sigma(\cdot)$ denotes a nonlinear activation function.

We parameterize the rectification process in residual form, $\Delta\mathbf{p}=\mathbf{p}_{\mathrm{clean}}-\mathbf{p}$ and $\mathbf{p}_{\mathrm{edit}}=\mathbf{p}+\Delta\mathbf{p}$, where $\Delta\mathbf{p}$ is the rectification vector and $\mathbf{p}_{\mathrm{edit}}$ is the edited preference representation. Then, the preference score is computed as $s_{\mathrm{pref}}=\sigma(\mathbf{w}_{\mathrm{pref}}^{\top}\mathbf{p}_{\mathrm{edit}}+b_{\mathrm{pref}})$, where $s_{\mathrm{pref}}$ is the predicted preference score, and $\mathbf{w}_{\mathrm{pref}}$ and $b_{\mathrm{pref}}$ are learnable parameters.

To prevent excessive modification of the original preference representation, we introduce two regularization terms:
\begin{equation}
\label{eq:mag_loss}
\mathcal{L}_{\mathrm{mag}}
=
\frac{1}{B}\sum_{i=1}^{B}
\left\|\Delta\mathbf{p}_{i}\right\|_{2}^{2}.
\end{equation}

\begin{equation}
\label{eq:dir_loss}
\mathcal{L}_{\mathrm{dir}}
=
\frac{1}{B}\sum_{i=1}^{B}
\left(
1-
\cos(\mathbf{p}_{\mathrm{edit},i},\mathbf{p}_{i})
\right).
\end{equation}
where $B$ is the mini-batch size, $\Delta\mathbf{p}_{i}$ is the residual rectification vector of the $i$-th sample, $\mathbf{p}_{\mathrm{edit},i}$ and $\mathbf{p}_{i}$ are its edited and original preference representations, and $\cos(\cdot,\cdot)$ denotes cosine similarity. $\mathcal{L}_{\mathrm{mag}}$ penalizes large residual edits, while $\mathcal{L}_{\mathrm{dir}}$ encourages the edited preference representation to preserve the semantic direction of the original one.

\subsection{LLM-Driven Semantic Anchoring}
\label{subsec:rpc}

To improve the semantic discriminability of the relevance representation, we introduce two fixed semantic prototypes, denoted by $\mathbf{r}_{\mathrm{pos}}$ and $\mathbf{r}_{\mathrm{neg}}$. The prototype $\mathbf{r}_{\mathrm{pos}}$ represents the semantic pole of relevant query-item pairs, while $\mathbf{r}_{\mathrm{neg}}$ represents the semantic pole of irrelevant query-item pairs. These prototypes are pre-encoded by BGE-base-en-v1.5, an external text encoder from the BGE family~\cite{xiao2024cpack}, and remain fixed during training.

Given the relevance representation $\mathbf{r}$, we project it into the prototype anchoring space by $\mathbf{r}_{\mathrm{proto}}=f_{\mathrm{proj}}(\mathbf{r})$, where $f_{\mathrm{proj}}(\cdot)$ is a learnable projection function and $\mathbf{r}_{\mathrm{proto}}$ is the projected relevance representation.
After $\ell_{2}$ normalization of both the projected relevance representation and the semantic prototypes, we compute the cosine similarities between $\mathbf{r}_{\mathrm{proto}}$ and the two semantic prototypes:
\begin{equation}
\label{eq:prompt_score_pos}
s_{\mathrm{pos}}
=
\cos(\mathbf{r}_{\mathrm{proto}},
\mathbf{r}_{\mathrm{pos}})
/
T_{\mathrm{proto}},
\end{equation}
\begin{equation}
\label{eq:prompt_score_neg}
s_{\mathrm{neg}}
=
\cos(\mathbf{r}_{\mathrm{proto}},
\mathbf{r}_{\mathrm{neg}})
/
T_{\mathrm{proto}},
\end{equation}
where $s_{\mathrm{pos}}$ and $s_{\mathrm{neg}}$ are the similarity scores with respect to the positive and negative relevance prototypes, respectively, and $T_{\mathrm{proto}}$ is the prototype temperature.

For the $i$-th sample, let $t_i\in\{+1,-1\}$ denote the relevance supervision label for prototype anchoring, where $+1$ and $-1$ indicates a relevant and irrelevant sample, respectively. Let $m_i\in\{0,1\}$ be a validity mask, where $m_i=1$ means that the relevance supervision is reliable and available, and $m_i=0$ means that the sample is excluded from this auxiliary loss. The prototype anchoring loss is defined as
\begin{equation}
\label{eq:prompt_loss}
\mathcal{L}_{\mathrm{pr}}
=
\frac{1}{\sum_{i=1}^{B}m_i+\varepsilon}
\sum_{i=1}^{B}
m_i
\max
\Big(
0,\,
\gamma
-
t_i\Delta s_i
\Big),
\end{equation}
where $\Delta s_i=s_{\mathrm{rel-pos}}^{i}-s_{\mathrm{rel-neg}}^{i}$ denotes the similarity gap between the positive and negative prototypes for the $i$-th sample, $\varepsilon$ is a small constant for numerical stability, and $\gamma$ is the margin hyperparameter. This loss aligns samples with their corresponding prototypes.

\subsection{Preference-Conditioned Evidence Routing}
\label{subsec:cer}

After obtaining the edited preference representation $\mathbf{p}_{\mathrm{edit}}$, we refine the relevance branch through preference-conditioned evidence routing. Let $\mathbf{q}$, $\mathbf{t}$, and $\mathbf{u}$ denote the encoded query representation, target item representation, and user representation, respectively. We construct six evidence tokens from the query, target item, user, and their pairwise interactions:
{\small
\begin{equation}
\label{eq:evidence_tokens_single}
\begin{aligned}
\mathbf{e}_{q} &= \tanh(\mathbf{W}_{q}\mathbf{q}),\;
\mathbf{e}_{t} = \tanh(\mathbf{W}_{t}\mathbf{t}),\;
\mathbf{e}_{u} = \tanh(\mathbf{W}_{u}\mathbf{u}),
\end{aligned}
\end{equation}
}
{\small
\begin{equation}
\label{eq:evidence_tokens_pairwise}
\begin{aligned}
\mathbf{e}_{qt} &= f_{qt}([\mathbf{q};\mathbf{t}]),\;
\mathbf{e}_{qu} = f_{qu}([\mathbf{q};\mathbf{u}]),\;
\mathbf{e}_{ut} = f_{ut}([\mathbf{u};\mathbf{t}]),
\end{aligned}
\end{equation}
}
\noindent where $\mathcal{E}=[\mathbf{e}_{q},\mathbf{e}_{t},\mathbf{e}_{u},\mathbf{e}_{qt},\mathbf{e}_{qu},\mathbf{e}_{ut}]$ denotes the ordered collection of evidence tokens. $\mathbf{W}_{q}$, $\mathbf{W}_{t}$, and $\mathbf{W}_{u}$ are learnable projection matrices, and $f_{qt}(\cdot)$, $f_{qu}(\cdot)$, and $f_{ut}(\cdot)$ are learnable interaction functions for query-item, query-user, and user-item evidence, respectively.

Using $\mathbf{p}_{\mathrm{edit}}$ as the routing condition, we compute the routing query $\mathbf{q}_{r}=\mathbf{W}_{q}^{r}\mathbf{p}_{\mathrm{edit}}$ and, for each evidence token $\mathbf{e}\in\mathcal{E}$, compute its key as $\mathbf{k}(\mathbf{e})=\mathbf{W}_{k}^{r}\mathbf{e}$. Here, $\mathbf{W}_{q}^{r}$ and $\mathbf{W}_{k}^{r}$ are learnable projection matrices.
The routing score and routing weight are computed for each evidence token $\mathbf{e}\in\mathcal{E}$ as
\begin{equation}
\label{eq:routing_score}
a(\mathbf{e})
=
\mathbf{w}_{s}^{\top}
\tanh(\mathbf{q}_{r}+\mathbf{k}(\mathbf{e})),
\end{equation}
\begin{equation}
\label{eq:routing_weight}
\alpha(\mathbf{e})
=
\frac{1}{
\sum_{\mathbf{x}\in\mathcal{E}}\exp(a(\mathbf{x})/\tau)
}
\cdot
\exp(a(\mathbf{e})/\tau),
\end{equation}
where $a(\mathbf{e})$ is the routing score of evidence token $\mathbf{e}$, $\alpha(\mathbf{e})$ is its normalized routing weight, $\mathbf{w}_{s}$ is a learnable scoring vector, $\tau>0$ is the routing temperature, and $\mathbf{x}$ indexes the evidence tokens in the denominator. The routed evidence summary is $\mathbf{c}_{\mathrm{route}}=\sum_{\mathbf{e}\in\mathcal{E}}\alpha(\mathbf{e})\mathbf{e}$, where $\mathbf{c}_{\mathrm{route}}$ is the preference-conditioned evidence representation.
We then construct an interaction-aware correction feature:
\begin{equation}
\label{eq:corr_feature}
\begin{aligned}
\mathbf{z}_{\mathrm{corr}}
=
[
&\mathbf{r};
\mathbf{c}_{\mathrm{route}};
\mathbf{r}\odot\mathbf{c}_{\mathrm{route}};
|\mathbf{r}-\mathbf{c}_{\mathrm{route}}|
],
\end{aligned}
\end{equation}
where $\mathbf{z}_{\mathrm{corr}}$ is the correction feature, $\odot$ denotes element-wise multiplication, and $|\cdot|$ denotes the element-wise absolute value.

A correction network predicts a relevance adjustment term $\delta_{\mathrm{rel}}=\mathrm{MLP}_{\mathrm{corr}}(\mathbf{z}_{\mathrm{corr}})$, where $\mathrm{MLP}_{\mathrm{corr}}(\cdot)$ is a multi-layer perceptron and $\delta_{\mathrm{rel}}$ is a scalar correction term in the relevance logit space. Let the original relevance score be $s_{\mathrm{rel}}=\sigma(\mathbf{w}_{\mathrm{rel}}^{\top}\mathbf{r}+b_{\mathrm{rel}})$, where $\mathbf{w}_{\mathrm{rel}}$ and $b_{\mathrm{rel}}$ are learnable parameters. We refine the relevance score in the logit space:
\begin{equation}
\label{eq:rel_refine}
s_{\mathrm{rel}}'
=
\sigma\left(
\operatorname{logit}(s_{\mathrm{rel}})
+
\delta_{\mathrm{rel}}
\right),
\end{equation}
where $s_{\mathrm{rel}}'$ is the refined relevance score and $\operatorname{logit}(x)=\log\frac{x}{1-x}$ denotes the inverse sigmoid function.

To regularize the evidence routing distribution, we introduce an entropy-based term:
\begin{equation}
\label{eq:attn_reg}
\mathcal{L}_{\mathrm{att}}
=
-\frac{1}{B}
\sum_{i=1}^{B}
\sum_{\mathbf{e}\in\mathcal{E}}
\alpha_i(\mathbf{e})
\log(\alpha_i(\mathbf{e})+\varepsilon),
\end{equation}
where $\alpha_i(\mathbf{e})$ is the routing weight of evidence token $\mathbf{e}\in\mathcal{E}$ for the $i$-th sample, $B$ is the mini-batch size, and $\varepsilon$ is a small constant for numerical stability. This term regularizes the entropy of the routing distribution and controls the selectivity of evidence aggregation.

\subsection{Final Prediction and Training Objective}
\label{subsec:loss}

The final behavior prediction is jointly determined by the preference score and the refined relevance score. Instead of imposing a fixed multiplicative assumption, we use an interaction function $g_{\theta}(\cdot,\cdot)$:
\begin{equation}
\label{eq:final_prediction}
\hat{y}_{i}
=
g_{\theta}
\left(
s_{\mathrm{pref},i},
s_{\mathrm{rel},i}'
\right),
\end{equation}
where $\hat{y}_{i}$ is the predicted behavior score of the $i$-th sample, $s_{\mathrm{pref},i}$ is its preference score, $s_{\mathrm{rel},i}'$ is its refined relevance score, and $g_{\theta}(\cdot,\cdot)$ is a learnable interaction function parameterized by $\theta$.

Given the ground-truth binary label $y_i\in\{0,1\}$, the main behavior prediction loss is the point-wise binary cross-entropy:
\begin{equation}
\label{eq:bce_loss}
\footnotesize
\mathcal{L}_{\mathrm{bce}}
=
-\frac{1}{B}
\sum_{i=1}^{B}
\left[
y_i\log(\hat{y}_i+\varepsilon)
+
(1-y_i)\log(1-\hat{y}_i+\varepsilon)
\right],
\end{equation}
where $B$ is the mini-batch size and $\varepsilon$ is a small constant for numerical stability.

The overall training objective combines the main prediction loss and all auxiliary losses:
\begin{equation}
\label{eq:overall_loss}
\footnotesize
\mathcal{L}
=
\mathcal{L}_{\mathrm{bce}}
+
\lambda_{\mathrm{mag}}\mathcal{L}_{\mathrm{mag}}
+
\lambda_{\mathrm{dir}}\mathcal{L}_{\mathrm{dir}}
+
\lambda_{\mathrm{pr}}\mathcal{L}_{\mathrm{pr}}
+
\lambda_{\mathrm{att}}\mathcal{L}_{\mathrm{att}}.
\end{equation}
where $\mathcal{L}_{\mathrm{bce}}$ is the main behavior prediction loss, $\mathcal{L}_{\mathrm{mag}}$ and $\mathcal{L}_{\mathrm{dir}}$ regularize preference rectification, $\mathcal{L}_{\mathrm{pr}}$ regularizes semantic prototype anchoring, and $\mathcal{L}_{\mathrm{att}}$ regularizes evidence routing. The coefficients $\lambda_{\mathrm{mag}}$, $\lambda_{\mathrm{dir}}$, $\lambda_{\mathrm{pr}}$, and $\lambda_{\mathrm{att}}$ are hyperparameters controlling the contribution of each auxiliary objective.

\section{Experiment}


\subsection{Experiment Setting}

\subsubsection{Experiment Environment.}

We implement the proposed method using PyTorch 2.7.0 and Python 3.12 on Ubuntu 22.04. All experiments are conducted with CUDA 12.8 on a single NVIDIA GeForce RTX 4090 GPU with 24 GB memory. The experimental machine is equipped with a 16-vCPU Intel(R) Xeon(R) Platinum 8352V CPU @ 2.10 GHz. We optimize the model using Adam~\cite{Kingma2015Adam}.

\subsubsection{Dataset.}

We evaluate PRISM on two public datasets, KuaiSAR~\cite{Sun2023KuaiSAR} and JD Search~\cite{Liu2023JDsearch}. KuaiSAR is a unified search and recommendation dataset released by the short-video platform Kuaishou. From this dataset, we isolate the video search logs to evaluate the effectiveness of PRISM in non-e-commerce scenarios. JD Search is constructed by the e-commerce platform Jingdong and represents a typical e-commerce scenario. The detailed statistics, including the number of users, items, queries, interactions, and search sessions, are listed in Table~\ref{tab:dataset_statistics}. For data splitting, the search sessions in the first 80\% of the time period are used as the training set, the middle 10\% as the validation set, and the remaining 10\% as the test set.

\subsubsection{Backbones \& Baselines.}

To evaluate the flexibility of our framework, we instantiate it with a variety of relevance and preference backbones for comparison. Specifically, the relevance backbones include \textit{DSSM}~\cite{Huang2013DSSM}, \textit{QEM}~\cite{Ai2017HEM}, and \textit{HEM}~\cite{Ai2017HEM}, where \textit{DSSM} and \textit{QEM} are ad-hoc relevance models, while \textit{HEM} is a personalized relevance model. For preference modeling, we adopt \textit{MLP}~\cite{Zhang2016MLP} and \textit{DCN}~\cite{Wang2017DCN}. In addition, we compare our method with several joint modeling baselines, including \textit{CLK}~\cite{Yao2021CLK}, \textit{NISE}~\cite{Huang2024NISE}, \textit{DCMT}~\cite{Zhu2023DCMT}, and \textit{PRINT}~\cite{Hong2024PRINT}.

\begin{table}[!t]
\caption{Dataset statistics.}
\label{tab:dataset_statistics}
\centering
\footnotesize
\setlength{\tabcolsep}{3pt}
\renewcommand{\arraystretch}{1.0}

\begin{tabular}{lccccc}
\toprule
Dataset & \# User & \# Item & \# Query & \# Interaction & \# Session \\
\midrule
KuaiSAR  & 25,877 & 2,012,476 & 175,849 & 3,171,231 & 267,608 \\
JDSearch & --     & 12,141,247 & 111,556 & 15,510,012 & 173,831 \\
\bottomrule
\end{tabular}
\end{table}

\subsubsection{Evaluation Metrics.}
Following previous studies~\cite{Guo2017DeepFM}, we adopt the Area Under the ROC Curve (AUC) and LogLoss as the primary evaluation metrics. In addition, to further assess ranking quality, we also report the Top-10 Hit Rate~(HR) and the Top-10 Normalized Discounted Cumulative Gain~(NDCG). Note that, for these two ranking-based metrics, we only consider sessions that contain positive samples.

\subsubsection{Implementation Details.}
Following UniSAR~\cite{Shi2024UniSAR}, we encode anonymous textual features in JDSearch and KuaiSAR, and encode user behavior sequences for all datasets. All textual and sparse feature representations are projected into a 64-dimensional embedding feature. The hidden dimensions of the final prediction network are set to $[64, 32]$. The routing temperature in Preference-Conditioned Evidence Routing is set to 1.0. To stabilize training, we initialize the last layer of the relevance correction network and the routing score projection to zero. The overall training objective consists of the main BCE loss, the relevance auxiliary term, the prompt alignment loss, and the regularization terms of the PR module. The weight of the main BCE loss is set to 1, the weight of the relevance auxiliary term is set to 0.001, and the weight of the prompt alignment loss is set to 0.1. For the PR module, the weights of magnitude regularization, direction regularization, and routing entropy regularization are set to $1\times10^{-4}$, $1\times10^{-3}$, and $1\times10^{-4}$, respectively. The weight decay is set to $1\times10^{-6}$. We adopt early stopping based on the validation performance, with the patience set to 2.

\begin{table*}[!t]
\centering
\caption{Overall comparison on KuaiSAR and JDSearch (PM = MLP).}
\label{tab:overall_mlp}
\scriptsize
\setlength{\tabcolsep}{4.6pt}
\renewcommand{\arraystretch}{1.08}
\begin{adjustbox}{max totalsize={\textwidth}{0.90\textheight},center}
\begin{tabular}{c c c | c c c c | c c c c}
\toprule
\multirow{2}{*}{PM} & \multirow{2}{*}{RM} & \multirow{2}{*}{JM}
& \multicolumn{4}{c|}{KuaiSAR}
& \multicolumn{4}{c}{JDSearch} \\
\cmidrule(lr){4-7}\cmidrule(lr){8-11}
& & & AUC $\uparrow$ & LogLoss $\downarrow$ & NDCG@10 $\uparrow$ & HR@10 $\uparrow$ & AUC $\uparrow$ & LogLoss $\downarrow$ & NDCG@10 $\uparrow$ & HR@10 $\uparrow$ \\
\midrule
\multirow{3}{*}{-}
& DSSM & Base & 0.57658 & 0.36103 & 0.47542 & 0.87159 & 0.67397 & 0.09710 & 0.26075 & 0.50674 \\
& QEM  & Base & 0.58398 & 0.35226 & 0.47782 & 0.87087 & 0.68705 & 0.09423 & 0.27128 & 0.51843 \\
& HEM  & Base & 0.62190 & 0.34482 & 0.47717 & 0.87217 & 0.68499 & 0.09426 & 0.27159 & 0.51738 \\
\midrule
\multirow{21}{*}{MLP}
& - & - & 0.62508 & 0.34536 & 0.48236 & 0.87465 & 0.68591 & 0.09419 & 0.27130 & 0.51778 \\
\cmidrule(lr){2-11}
& \multirow{6}{*}{DSSM}
& Base  & 0.62741 & 0.34473 & 0.48120 & 0.87362 & \underline{0.69311} & \textbf{0.09394} & \underline{0.27264} & \underline{0.52756} \\
& & CLK   & 0.62728 & 0.34474 & 0.48110 & 0.87351 & 0.69199 & \underline{0.09405} & 0.27188 & 0.52611 \\
& & NISE  & 0.62744 & 0.34473 & 0.48126 & \underline{0.87365} & 0.69199 & \underline{0.09405} & 0.27188 & 0.52611 \\
& & PRINT & 0.62490 & \underline{0.34421} & 0.47705 & 0.87100 & 0.68855 & 0.09415 & 0.26946 & 0.51527 \\
& & DRP   & \underline{0.62877} & 0.34429 & \underline{0.48234} & 0.87342 & 0.69183 & 0.09406 & 0.27022 & 0.52468 \\
& & PRISM   & \textbf{0.63422}$^{*}$ & \textbf{0.34199}$^{*}$ & \textbf{0.48775}$^{*}$ & \textbf{0.87571}$^{*}$ & \textbf{0.70876}$^{*}$ & 0.09640 & \textbf{0.28604}$^{*}$ & \textbf{0.56010}$^{*}$ \\
\cmidrule(lr){2-11}
& \multirow{7}{*}{QEM}
& Base  & 0.62697 & 0.34422 & 0.47972 & 0.87267 & 0.69347 & 0.09540 & 0.26846 & 0.52970 \\
& & CLK   & 0.62681 & 0.34423 & 0.47960 & 0.87286 & 0.69176 & 0.09531 & 0.27046 & 0.52970 \\
& & NISE  & 0.62692 & 0.34422 & 0.47983 & 0.87318 & 0.69176 & 0.09531 & 0.27046 & 0.52970 \\
& & DCMT  & 0.62735 & 0.34394 & 0.47944 & 0.87169 & 0.69094 & \underline{0.09492} & \underline{0.27084} & 0.52649 \\
& & PRINT & 0.62755 & 0.34415 & 0.47866 & 0.87261 & 0.69075 & \textbf{0.09452} & 0.26971 & 0.52194 \\
& & DRP   & \underline{0.62839} & \underline{0.34386} & \underline{0.48145} & \underline{0.87409} & \underline{0.69707} & 0.09742 & 0.26992 & \underline{0.53866} \\
& & PRISM   & \textbf{0.63307}$^{*}$ & \textbf{0.34312}$^{*}$ & \textbf{0.48851}$^{*}$ & \textbf{0.87695}$^{*}$ & \textbf{0.70298}$^{*}$ & 0.10054 & \textbf{0.28136}$^{*}$ & \textbf{0.55394}$^{*}$ \\
\cmidrule(lr){2-11}
& \multirow{7}{*}{HEM}
& Base  & 0.62835 & 0.34399 & 0.47953 & 0.87254 & 0.69096 & 0.09559 & 0.27042 & 0.52924 \\
& & CLK   & 0.62839 & 0.34397 & 0.47938 & 0.87214 & 0.69378 & 0.09652 & 0.26987 & 0.53727 \\
& & NISE  & 0.62841 & 0.34399 & 0.47934 & 0.87262 & 0.69378 & 0.09652 & 0.26987 & 0.53727 \\
& & DCMT  & 0.62858 & \underline{0.34369} & 0.47931 & 0.87314 & 0.69063 & \textbf{0.09449} & \underline{0.27202} & 0.52354 \\
& & PRINT & 0.62823 & 0.34419 & 0.47788 & 0.87312 & 0.68940 & \underline{0.09458} & 0.27050 & 0.52129 \\
& & DRP   & \underline{0.62884} & 0.34376 & \underline{0.48268} & \underline{0.87383} & \underline{0.69591} & 0.09770 & 0.27008 & \underline{0.53741} \\
& & PRISM   & \textbf{0.63237}$^{*}$ & \textbf{0.34323} & \textbf{0.48708}$^{*}$ & \textbf{0.87630}$^{*}$ & \textbf{0.70339}$^{*}$ & 0.10189 & \textbf{0.28244}$^{*}$ & \textbf{0.55880}$^{*}$ \\
\bottomrule
\end{tabular}
\end{adjustbox}
\end{table*}

\begin{table*}[!t]
\centering
\caption{Overall comparison on KuaiSAR and JDSearch (PM = DCN).}
\label{tab:overall_dcn}
\scriptsize
\setlength{\tabcolsep}{4.6pt}
\renewcommand{\arraystretch}{1.08}
\begin{adjustbox}{max totalsize={\textwidth}{0.90\textheight},center}
\begin{tabular}{c c c | c c c c | c c c c}
\toprule
\multirow{2}{*}{PM} & \multirow{2}{*}{RM} & \multirow{2}{*}{JM}
& \multicolumn{4}{c|}{KuaiSAR}
& \multicolumn{4}{c}{JDSearch} \\
\cmidrule(lr){4-7}\cmidrule(lr){8-11}
& & & AUC $\uparrow$ & LogLoss $\downarrow$ & NDCG@10 $\uparrow$ & HR@10 $\uparrow$ & AUC $\uparrow$ & LogLoss $\downarrow$ & NDCG@10 $\uparrow$ & HR@10 $\uparrow$ \\
\midrule
\multirow{3}{*}{-}
& DSSM & Base & 0.57658 & 0.36103 & 0.47542 & 0.87159 & 0.67397 & 0.09710 & 0.26075 & 0.50674 \\
& QEM  & Base & 0.58398 & 0.35226 & 0.47782 & 0.87087 & 0.68705 & 0.09423 & 0.27128 & 0.51843 \\
& HEM  & Base & 0.62190 & 0.34482 & 0.47717 & 0.87217 & 0.68499 & 0.09426 & 0.27159 & 0.51738 \\
\midrule
\multirow{20}{*}{DCN}
& \multirow{6}{*}{DSSM}
& Base  & 0.62515 & 0.34485 & 0.48062 & 0.87465 & 0.69628 & \underline{0.09372} & \underline{0.27558} & 0.52935 \\
& & CLK   & 0.62509 & 0.34484 & 0.48059 & \underline{0.87497} & 0.69507 & 0.09382 & 0.27475 & 0.52838 \\
& & NISE  & 0.62515 & 0.34484 & 0.48064 & 0.87482 & 0.69507 & 0.09382 & 0.27475 & 0.52838 \\
& & PRINT & 0.62218 & 0.34482 & 0.47990 & 0.87275 & 0.69484 & \textbf{0.09370} & 0.27386 & 0.52486 \\
& & DRP   & \underline{0.62758} & \underline{0.34386} & \underline{0.48153} & 0.87428 & \underline{0.70526} & 0.09708 & 0.27281 & \underline{0.53797} \\
& & PRISM   & \textbf{0.63078}$^{*}$ & \textbf{0.34332} & \textbf{0.48621}$^{*}$ & \textbf{0.87557} & \textbf{0.71412}$^{*}$ & 0.09523 & \textbf{0.28950}$^{*}$ & \textbf{0.56552}$^{*}$ \\
\cmidrule(lr){2-11}

& \multirow{7}{*}{QEM}
& Base  & 0.62670 & 0.34428 & 0.48014 & 0.87231 & 0.69919 & 0.09619 & 0.28226 & \underline{0.54984} \\
& & CLK   & 0.62665 & 0.34427 & 0.48006 & 0.87300 & 0.69807 & \textbf{0.09478} & \underline{0.28339} & 0.54200 \\
& & NISE  & 0.62678 & 0.34427 & 0.48009 & 0.87274 & 0.69807 & \textbf{0.09478} & \underline{0.28339} & 0.54200 \\
& & DCMT  & 0.62613 & \underline{0.34408} & \underline{0.48217} & \underline{0.87376} & 0.70207 & 0.09606 & 0.28288 & 0.54803 \\
& & PRINT & 0.62660 & 0.34414 & 0.48062 & 0.87298 & 0.69996 & \underline{0.09589} & 0.27690 & 0.54071 \\
& & DRP   & \underline{0.62948} & \textbf{0.34389} & 0.48141 & 0.87367 & \underline{0.70297} & 0.09797 & 0.27070 & 0.53274 \\
& & PRISM   & \textbf{0.63115} & 0.34421 & \textbf{0.48704}$^{*}$ & \textbf{0.87512} & \textbf{0.71010}$^{*}$ & 0.09619 & \textbf{0.28406} & \textbf{0.55636}$^{*}$ \\
\cmidrule(lr){2-11}

& \multirow{7}{*}{HEM}
& Base  & 0.62802 & 0.34398 & 0.48133 & 0.87359 & 0.70058 & \underline{0.09596} & 0.28335 & \underline{0.55054} \\
& & CLK   & 0.62827 & \underline{0.34393} & 0.48111 & 0.87282 & 0.70034 & 0.09601 & \textbf{0.28434} & 0.55005 \\
& & NISE  & 0.62805 & 0.34398 & 0.48123 & 0.87359 & 0.70034 & 0.09601 & \textbf{0.28434} & 0.55005 \\
& & DCMT  & 0.62682 & 0.34419 & 0.48128 & 0.87364 & 0.70021 & 0.09634 & 0.28322 & 0.54746 \\
& & PRINT & 0.62591 & 0.34429 & 0.48208 & 0.87414 & 0.69915 & 0.09623 & 0.28116 & 0.54457 \\
& & DRP   & \underline{0.62900} & 0.34409 & \underline{0.48284} & \underline{0.87482} & \underline{0.70171} & 0.09764 & 0.26949 & 0.53027 \\
& & PRISM   & \textbf{0.63335}$^{*}$ & \textbf{0.34290}$^{*}$ & \textbf{0.48862}$^{*}$ & \textbf{0.87685} & \textbf{0.71216}$^{*}$ & \textbf{0.09564} & \underline{0.28384} & \textbf{0.55694}$^{*}$ \\
\bottomrule
\end{tabular}
\end{adjustbox}
\end{table*}

\subsection{Overall Performance}
We evaluate PRISM under various relevance and preference backbone combinations and compare it with existing methods for improving joint modeling. 
Each experiment is repeated five times with different random seeds for robustness. 
We report the mean performance over all runs, rounded to four decimal places. 
We further perform one-tailed $t$-tests to verify whether the best method significantly outperforms the runner-up. 
The best result is shown in bold, and the second-best result is underlined.

The overall comparison results under MLP and DCN as preference models are reported in Table~\ref{tab:overall_mlp} and Table~\ref{tab:overall_dcn}, respectively. 
Here, ``PM'' denotes the preference model, ``RM'' denotes the relevance model, and ``JM'' denotes the method applied to enhance the joint modeling framework, while ``-'' indicates that no such component is used. 
For instance, methods with both ``PM'' and ``JM'' set to ``-'' correspond to relevance-only models.

Overall, joint modeling methods outperform relevance-only and preference-only variants. 
This result indicates that combining relevance and preference signals is critical for better user intent modeling and improved ranking performance. 
It also confirms the necessity of designing effective unified joint modeling frameworks.

\begin{table*}[t]
\caption{Ablation results on KuaiSAR with MLP as the preference encoder and different relevance encoders.}
\label{tab:ablation}
\centering
\scriptsize
\setlength{\tabcolsep}{3.0pt}
\renewcommand{\arraystretch}{1.08}
\begin{tabular}{llcccrrrr}
\toprule
\textbf{PM} 
& \textbf{RM} 
& \makecell{\textbf{LLM-Driven}\\\textbf{Semantic Anchoring}}
& \makecell{\textbf{Preference-Conditioned}\\\textbf{Evidence Routing}}
& \makecell{\textbf{Preference}\\\textbf{Rectification }}
& \textbf{AUC} $\uparrow$ 
& \textbf{LogLoss} $\downarrow$ 
& \textbf{NDCG@10} $\uparrow$ 
& \textbf{HR@10} $\uparrow$ \\
\midrule
\multirow{18}{*}{MLP}
& \multirow{6}{*}{DSSM}
& $\times$ & $\times$ & $\times$ 
& 0.63005 & 0.34458 & 0.48612 & 0.87497 \\
& & $\checkmark$ & $\times$ & $\times$
& 0.59127 & 0.34958 & 0.48270 & 0.87293 \\
& & $\times$ & $\checkmark$ & $\checkmark$
& 0.63148 & 0.34362 & 0.48691 & 0.87466 \\
& & $\checkmark$ & $\times$ & $\checkmark$
& 0.63086 & 0.34404 & 0.48677 & 0.87514 \\
& & $\checkmark$ & $\checkmark$ & $\times$
& 0.58994 & 0.34955 & 0.48175 & 0.87297 \\
& & $\checkmark$ & $\checkmark$ & $\checkmark$
& \textbf{0.63422} & \textbf{0.34199} & \textbf{0.48775} & \textbf{0.87571} \\
\cmidrule(lr){2-9}

& \multirow{6}{*}{HEM}
& $\times$ & $\times$ & $\times$
& 0.63017 & 0.34375 & 0.48610 & 0.87422 \\
& & $\checkmark$ & $\times$ & $\times$
& 0.63014 & 0.34413 & 0.48442 & 0.87373 \\
& &$\times$ & $\checkmark$ & $\checkmark$
& 0.62997 & 0.34372 & 0.48515 & 0.87406 \\
& & $\checkmark$ & $\times$ & $\checkmark$
& 0.63048 & 0.34370 & 0.48525 & 0.87390 \\
& & $\checkmark$ & $\checkmark$ & $\times$
& 0.63097 & 0.34372 & 0.48492 & 0.87378 \\
& & $\checkmark$ & $\checkmark$ & $\checkmark$
& \textbf{0.63237} & \textbf{0.34323} & \textbf{0.48708} & \textbf{0.87630} \\
\cmidrule(lr){2-9}

& \multirow{6}{*}{QEM}
& $\times$ & $\times$ & $\times$
& 0.63014 & 0.34361 & 0.48547 & 0.87362 \\
& & $\checkmark$ & $\times$ & $\times$
& 0.58799 & 0.34972 & 0.48329 & 0.87303 \\
& & $\times$ & $\checkmark$ & $\checkmark$
& 0.62960 & 0.34365 & 0.48617 & 0.87488 \\
& & $\checkmark$ & $\times$ & $\checkmark$
& 0.62982 & 0.34414 & 0.48428 & 0.87328 \\
& & $\checkmark$ & $\checkmark$ & $\times$
& 0.58854 & 0.34879 & 0.48425 & 0.87376 \\
& & $\checkmark$ & $\checkmark$ & $\checkmark$
& \textbf{0.63307} & \textbf{0.34312} & \textbf{0.48851} & \textbf{0.87695} \\
\bottomrule
\end{tabular}
\end{table*}

\subsection{Objective Ablation}
\label{sec:ablation}

We conduct ablation studies on KuaiSAR using MLP as the preference encoder and three relevance encoders, i.e., DSSM, QEM, and HEM. Each configuration is repeated five times, and we report the averaged performance on the main test set in terms of AUC, Logloss, NDCG@10, and HR@10.
As shown in Table~\ref{tab:ablation}, PRISM consistently achieves the best overall performance across all three relevance encoders, demonstrating the effectiveness of the complete framework. Removing any major component generally leads to performance degradation, indicating that semantic anchoring, preference-conditioned evidence routing, and preference rectification are complementary to each other. In particular, the variants without Preference Rectification suffer noticeable performance drops, especially under DSSM and QEM, suggesting that explicitly rectifying preference representations is important for preventing noisy preference signals from harming downstream prediction. The results also show that LLM-driven semantic anchoring is most effective when combined with the other components, rather than being used in isolation. Overall, these findings confirm that the performance gains come from the integrated design of PRISM rather than from a single backbone or isolated module.

\begin{table*}[!t]
\caption{Loss ablation results on KuaiSAR under the QEM + MLP setting.}
\label{tab:loss_ablation}
\centering
\scriptsize
\begin{tabular}{ccccc rrrr}
\toprule
$\mathcal{L}_{\mathrm{bce}}$
& $\mathcal{L}_{\mathrm{mag}}$
& $\mathcal{L}_{\mathrm{dir}}$
& $\mathcal{L}_{\mathrm{pr}}$
& $\mathcal{L}_{\mathrm{att}}$
& \textbf{AUC} $\uparrow$
& \textbf{LogLoss} $\downarrow$
& \textbf{NDCG@10} $\uparrow$
& \textbf{HR@10} $\uparrow$ \\
\midrule
$\checkmark$ & $\times$ & $\times$ & $\times$ & $\times$
& 0.62916 & 0.34416 & 0.48521 & 0.87409 \\
$\checkmark$ & $\times$ & $\times$ & $\checkmark$ & $\times$
& 0.62905 & 0.34416 & 0.48527 & 0.87422 \\
$\checkmark$ & $\times$ & $\times$ & $\checkmark$ & $\checkmark$
& 0.62888 & 0.34420 & 0.48519 & 0.87433 \\
$\checkmark$ & $\checkmark$ & $\checkmark$ & $\times$ & $\checkmark$
& 0.62922 & 0.34409 & 0.48481 & 0.87335 \\
$\checkmark$ & $\checkmark$ & $\checkmark$ & $\checkmark$ & $\times$
& 0.62891 & 0.34412 & 0.48527 & 0.87396 \\
$\checkmark$ & $\checkmark$ & $\checkmark$ & $\checkmark$ & $\checkmark$
& \textbf{0.63307} & \textbf{0.34312} & \textbf{0.48851} & \textbf{0.87695} \\
\bottomrule
\end{tabular}
\end{table*}

\subsection{Loss Ablation Study}
\label{sec:loss_ablation}

We further study the contribution of different objective terms under the QEM + MLP setting on KuaiSAR, and the results are shown in Table~\ref{tab:loss_ablation}. It can be seen that the full objective consistently achieves the best overall performance. Removing the prompt-based supervision leads to the most noticeable degradation, indicating that it provides important complementary semantic guidance for ranking. Removing the editing losses or the routing regularization also weakens the results, suggesting that these terms help stabilize and refine the preference editing process. When all edit-related regularization terms are removed, the model still underperforms the full objective, showing that the gain comes not only from the architecture itself but also from the training objective imposed on editing and routing. Finally, using only the main BCE loss performs worse than the full objective, further confirming that auxiliary supervision is beneficial beyond the BCE loss alone.

\subsection{Performance on Frequency-based Subsets}
\label{sec:freq_subset}

To further evaluate robustness under different frequency regimes, we partition the test set into four subsets: \textit{ihot}, \textit{itail}, \textit{qhot}, and \textit{qtail}, corresponding to item-hot, item-tail, query-hot, and query-tail impressions, respectively.
Table~\ref{tab:subset_results} shows that our method consistently achieves the best performance on all four subsets. In particular, it yields the highest AUC, NDCG@10, and HR@10, together with the lowest Logloss, across both item-side and query-side hot/tail splits. The gains are especially evident on the query-based subsets, indicating that the proposed design better handles heterogeneous query distributions. Overall, these results demonstrate that our method not only improves overall ranking quality, but also generalizes robustly across both hot and tail regimes.
\begin{table}[t]
\caption{Performance comparison on frequency-based subsets of KuaiSAR under PM = DCN and RM = HEM.}
\label{tab:subset_results}
\centering
\footnotesize
\setlength{\tabcolsep}{2.6pt}
\renewcommand{\arraystretch}{1.0}

\begin{tabular}{llrrrr}
\toprule
Subset & Method & AUC$\uparrow$ & LogLoss$\downarrow$ & NDCG@10$\uparrow$ & HR@10$\uparrow$ \\
\midrule

\multirow{7}{*}{ihot}
& Base          & 0.69643 & 0.32989 & 0.67689 & 0.95652 \\
& PRINT         & 0.69706 & 0.32981 & 0.68000 & 0.95817 \\
& DRP           & 0.69590 & 0.33023 & 0.67874 & 0.95969 \\
& CLK           & 0.69624 & 0.32995 & 0.67664 & 0.95623 \\
& NISE          & 0.69643 & 0.32988 & 0.67686 & 0.95764 \\
& DCMT          & 0.69591 & 0.33001 & 0.67909 & 0.95837 \\
& PRISM         & \textbf{0.70369} & \textbf{0.32657} & \textbf{0.68472} & \textbf{0.96086} \\
\cmidrule(lr){1-6}

\multirow{7}{*}{itail}
& Base          & 0.60965 & 0.34794 & 0.49740 & 0.88200 \\
& PRINT         & 0.60693 & 0.34836 & 0.49631 & 0.88245 \\
& DRP           & 0.61124 & 0.34799 & 0.49837 & 0.88291 \\
& CLK           & 0.61000 & 0.34786 & 0.49743 & 0.88144 \\
& NISE          & 0.60971 & 0.34795 & 0.49726 & 0.88168 \\
& DCMT          & 0.60872 & 0.34818 & 0.49686 & 0.88248 \\
& PRISM         & \textbf{0.61277} & \textbf{0.34749} & \textbf{0.50110} & \textbf{0.88317} \\
\cmidrule(lr){1-6}

\multirow{7}{*}{qhot}
& Base          & 0.67059 & 0.32007 & 0.50019 & 0.87265 \\
& PRINT         & 0.66949 & 0.32003 & 0.50101 & 0.87257 \\
& DRP           & 0.66708 & 0.32080 & 0.49986 & 0.87580 \\
& CLK           & 0.67022 & 0.32020 & 0.49842 & 0.87192 \\
& NISE          & 0.67046 & 0.32007 & 0.49942 & 0.87265 \\
& DCMT          & 0.66752 & 0.32046 & 0.50062 & 0.87316 \\
& PRISM         & \textbf{0.67385} & \textbf{0.31766} & \textbf{0.50861} & \textbf{0.87814} \\
\cmidrule(lr){1-6}

\multirow{7}{*}{qtail}
& Base          & 0.61646 & 0.35003 & 0.47735 & 0.87379 \\
& PRINT         & 0.61411 & 0.35043 & 0.47747 & 0.87368 \\
& DRP           & 0.61876 & 0.34999 & 0.47924 & 0.87463 \\
& CLK           & 0.61688 & 0.34994 & 0.47745 & 0.87302 \\
& NISE          & 0.61654 & 0.35004 & 0.47740 & 0.87379 \\
& DCMT          & 0.61603 & 0.35020 & 0.47720 & 0.87374 \\
& PRISM         & \textbf{0.62248} & \textbf{0.34929} & \textbf{0.48440} & \textbf{0.87657} \\
\bottomrule
\end{tabular}
\end{table}

\begin{table*}[!t]
\caption{Effectiveness of Preference Rectification under the DSSM/QEM + MLP setting on KuaiSAR.}
\label{tab:pr_effectiveness}
\centering
\begin{tabular}{llrrrr}
\toprule
\textbf{RM}
& \textbf{Method} 
& \textbf{AUC} $\uparrow$ 
& \textbf{LogLoss} $\downarrow$ 
& \textbf{NDCG@10} $\uparrow$ 
& \textbf{HR@10} $\uparrow$ \\
\midrule
\multirow{4}{*}{DSSM}
& DRP-LowRankOrth
& 0.63129 & 0.34343 & 0.48762 & 0.87452 \\
& Simple Attention 
& 0.59071 & 0.34901 & 0.48447 & 0.87445 \\
& MLP Fusion 
& 0.63110 & 0.34328 & 0.48760 & 0.87602 \\
& \textbf{Preference Rectification}
& \textbf{0.63422} & \textbf{0.34199} & \textbf{0.48775} & 0.87571 \\
\cmidrule(lr){1-6}
\multirow{4}{*}{QEM}
& DRP-LowRankOrth
& 0.63197 & 0.34313 & 0.48809 & 0.87382 \\
& Simple Attention 
& 0.59027 & 0.34854 & 0.48794 & 0.87470 \\
& MLP Fusion 
& 0.63238 & 0.34320 & 0.48840 & 0.87482 \\
& \textbf{Preference Rectification}
& \textbf{0.63307} & \textbf{0.34312} & \textbf{0.48851} & \textbf{0.87695} \\
\bottomrule
\end{tabular}
\end{table*}

\subsection{Effectiveness of Preference Rectification}
\label{sec:preference_rectification}

To further verify the effectiveness of the proposed Preference Rectification module, we conduct an additional comparison based on the original DSSM/QEM + MLP setting on KuaiSAR. Specifically, we keep the backbone setting unchanged and replace the Preference Rectification module with several alternative methods for processing preference representations and relevance representations. The compared methods include MLP Fusion, which uses a feed-forward network to fuse the two representations; Simple Attention, which applies a lightweight attention mechanism for representation interaction; and DRP-LowRankOrth, which adopts the low-rank orthogonal decomposition strategy from DRP~\cite{Wang2025DRP}. The results are reported in Table~\ref{tab:pr_effectiveness}.
It can be noticed that replacing the proposed Preference Rectification module with other representation processing strategies generally leads to inferior performance. Compared with DRP-LowRankOrth~\cite{Wang2025DRP}, Preference Rectification achieves better overall results, indicating that the proposed module is more effective than low-rank orthogonal decomposition in modeling the interaction between preference representations and relevance representations.

In addition, Simple Attention shows a substantial performance drop, suggesting that simple attention-based fusion is insufficient to capture the structured dependencies required for preference rectification. Compared with these alternative designs, Preference Rectification provides more stable and effective representation refinement across different relevance models, demonstrating the effectiveness of the proposed module for downstream prediction.

\subsection{Visualization Analysis}

\begin{figure}[t]
    \centering
    \includegraphics[width=0.85\linewidth]{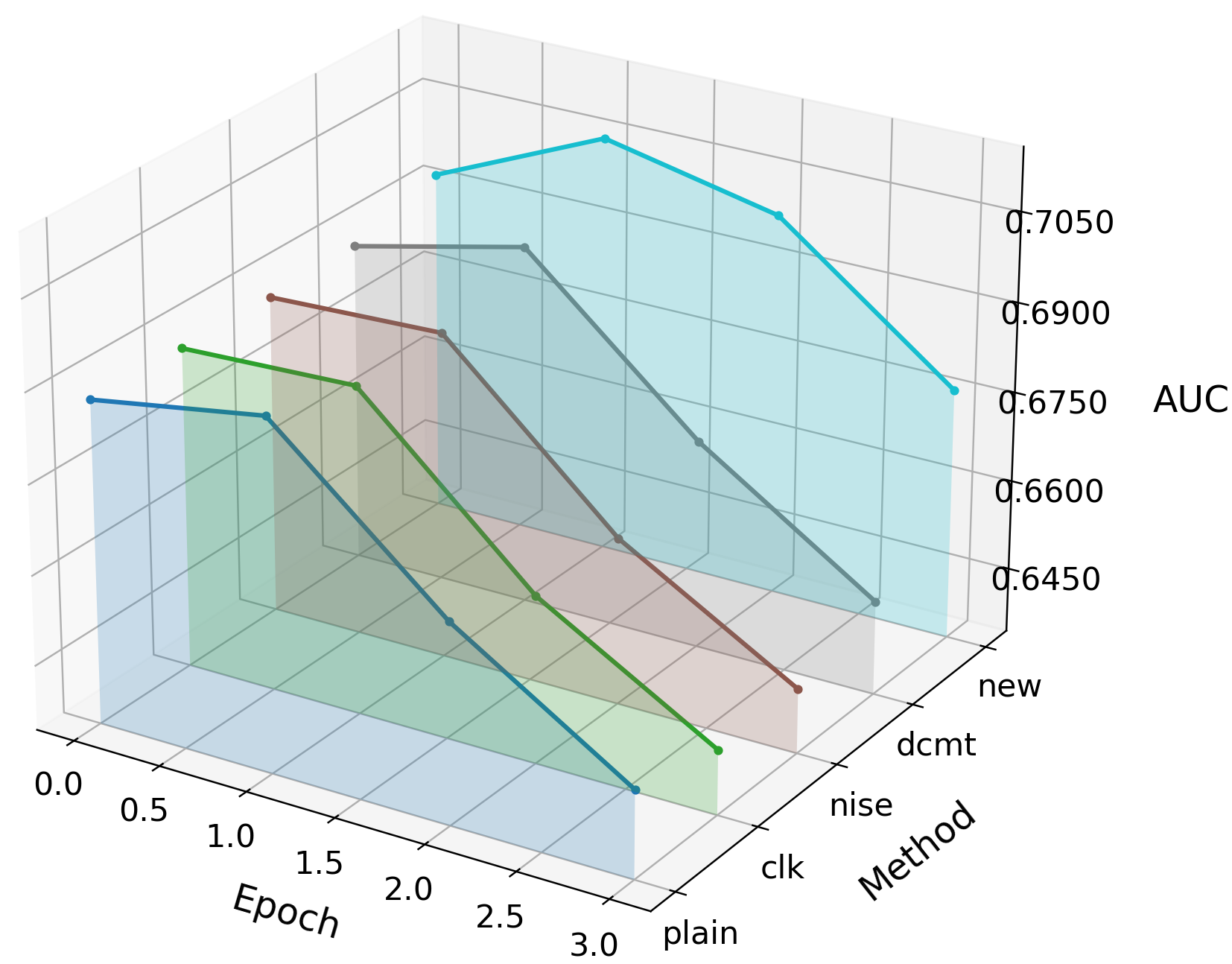}
    \caption{Validation AUC across training epochs for the proposed method and multiple baselines.}
    \label{fig:wall_epoch}
\end{figure}

\paragraph{Validation AUC across training epochs.}
Figure~\ref{fig:wall_epoch} compares the validation AUC trajectories of different methods across training epochs. 
Compared with the baseline methods, including \textit{plain}, \textit{clk}, \textit{nise}, and \textit{dcmt}, the proposed method consistently achieves superior validation AUC throughout the optimization process and exhibits a noticeably smoother performance curve. 
Although several baselines obtain moderate gains during the early stages of training, their validation performance deteriorates more substantially as training proceeds, suggesting that these methods provide limited structural regularization on the learned representation space and are therefore less effective at maintaining a stable discriminative structure during optimization. 
In contrast, our method remains in a higher AUC regime over all epochs and shows a smaller degree of late-stage degradation. 
This observation indicates that the introduced semantic constraints improve not only the attainable performance ceiling but also the stability of the training dynamics. 
From a representation learning perspective, the LLM-driven semantic constraints act as a structured prior that continuously guides the relevance representation toward a more discriminative and ranking-consistent geometry, thereby yielding more robust generalization behavior.

\begin{figure}[t]
    \centering
    \subfloat[]{
        \includegraphics[width=0.8\linewidth]{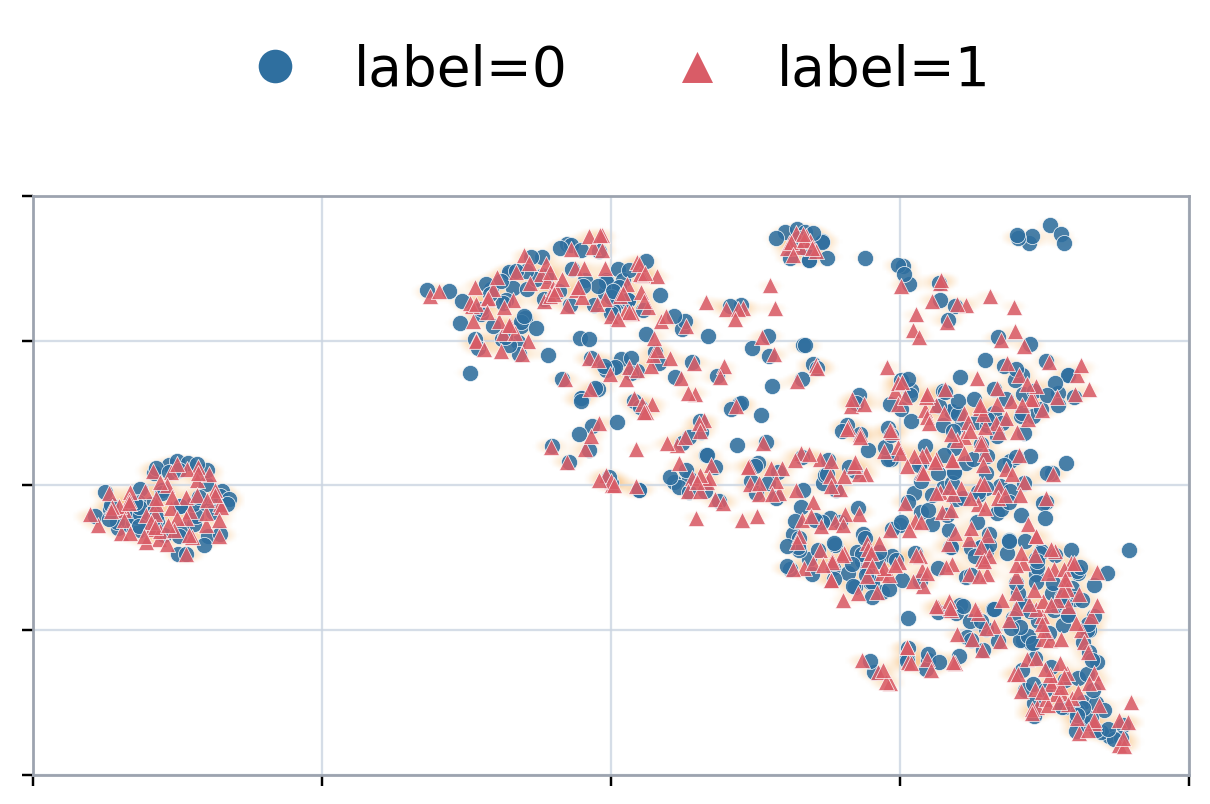}
        \label{fig:rel_tsne_a}
    }

    \vspace{1mm}

    \subfloat[]{
        \includegraphics[width=0.8\linewidth]{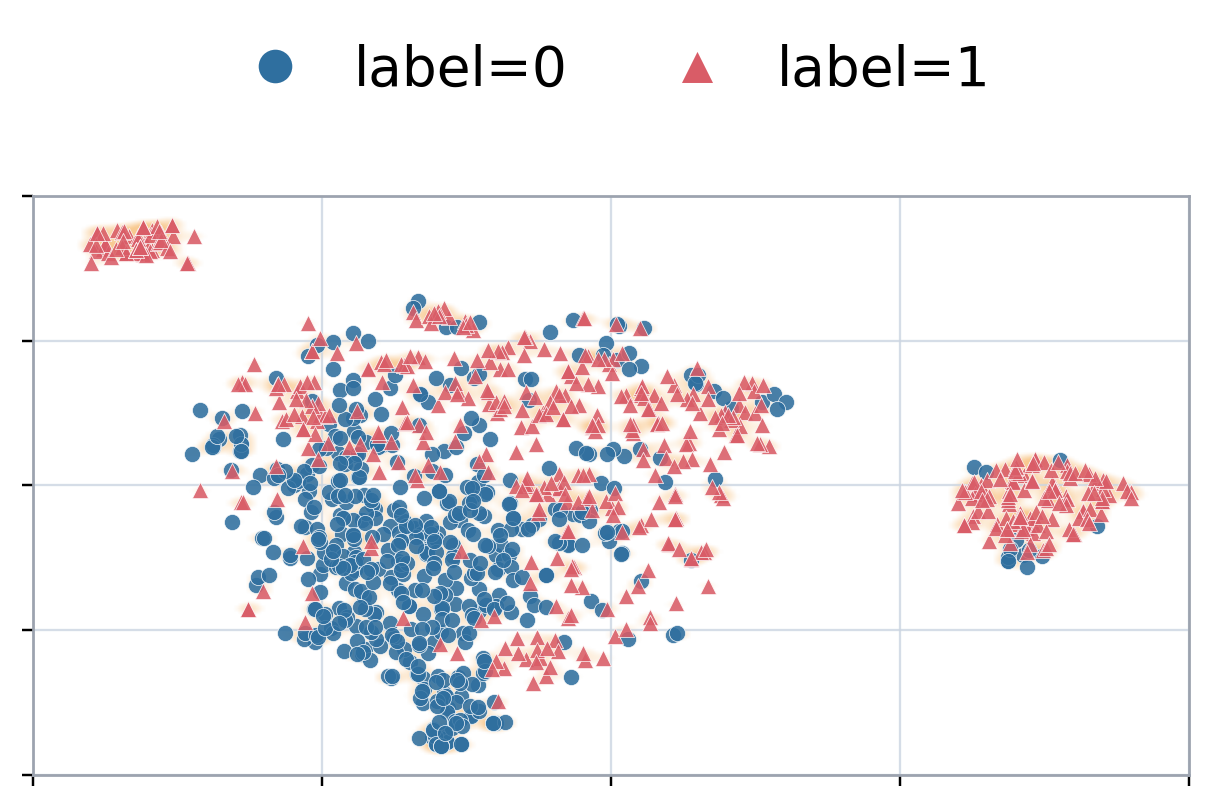}
        \label{fig:rel_tsne_b}
    }

    \caption{t-SNE visualization of the learned relevance representations, where label 0 denotes irrelevant samples and label 1 denotes relevant samples.}
    \label{fig:kuaisar_tsne}
\end{figure}

\paragraph{t-SNE visualization of the learned representation space.}
The t-SNE visualization in Figure~\ref{fig:kuaisar_tsne} provides further evidence on why the proposed method leads to better ranking performance. 
Without LLM-driven semantic constraints, positive and negative instances exhibit substantial overlap in the projected space, and the class boundary remains ambiguous. 
This suggests that, although the model is able to capture task-relevant signals to some extent, the resulting internal representation lacks a well-organized semantic structure. 
After introducing semantic constraints, the projected embeddings display a markedly improved separation pattern: samples from the same class become more compact, inter-class margins increase, and the class centroids are more clearly separated. 
These results imply that the semantic constraints do not merely affect the final prediction scores, but fundamentally reshape the geometry of the intermediate relevance representation. 
As a consequence, the learned space achieves better intra-class compactness and inter-class separability, which offers a plausible explanation for the consistent AUC gains observed during training. 
Overall, the visualization results support the claim that LLM-driven semantic constraints enhance representation quality by injecting semantically meaningful structural bias into the optimization process.




\section{Conclusion}

In this paper, we proposed PRISM, a unified preference--relevance interaction semantic modeling framework for search behavior prediction. PRISM addresses the entanglement between user preference and item relevance by introducing three complementary components: preference rectification, LLM-driven semantic anchoring, and preference-conditioned evidence routing. The preference rectification module suppresses relevance-induced interference in preference representations, the semantic anchoring module calibrates relevance representations with external semantic priors, and the evidence routing module adaptively aggregates heterogeneous evidence under the edited preference state. Extensive experiments on KuaiSAR and JDSearch demonstrate that PRISM consistently improves ranking performance over strong baselines across different backbone combinations. Further ablation studies and visualization analyzes verify the effectiveness of each component and show that PRISM learns more discriminative and semantically structured representations. In future work, we will explore better probability calibration and extend the framework to more complex multi-behavior search scenarios.

 

\bibliographystyle{IEEEtran}
\bibliography{TKDE/main}











\begin{IEEEbiography}
[{\includegraphics[width=1in,height=1.25in,clip,keepaspectratio]
{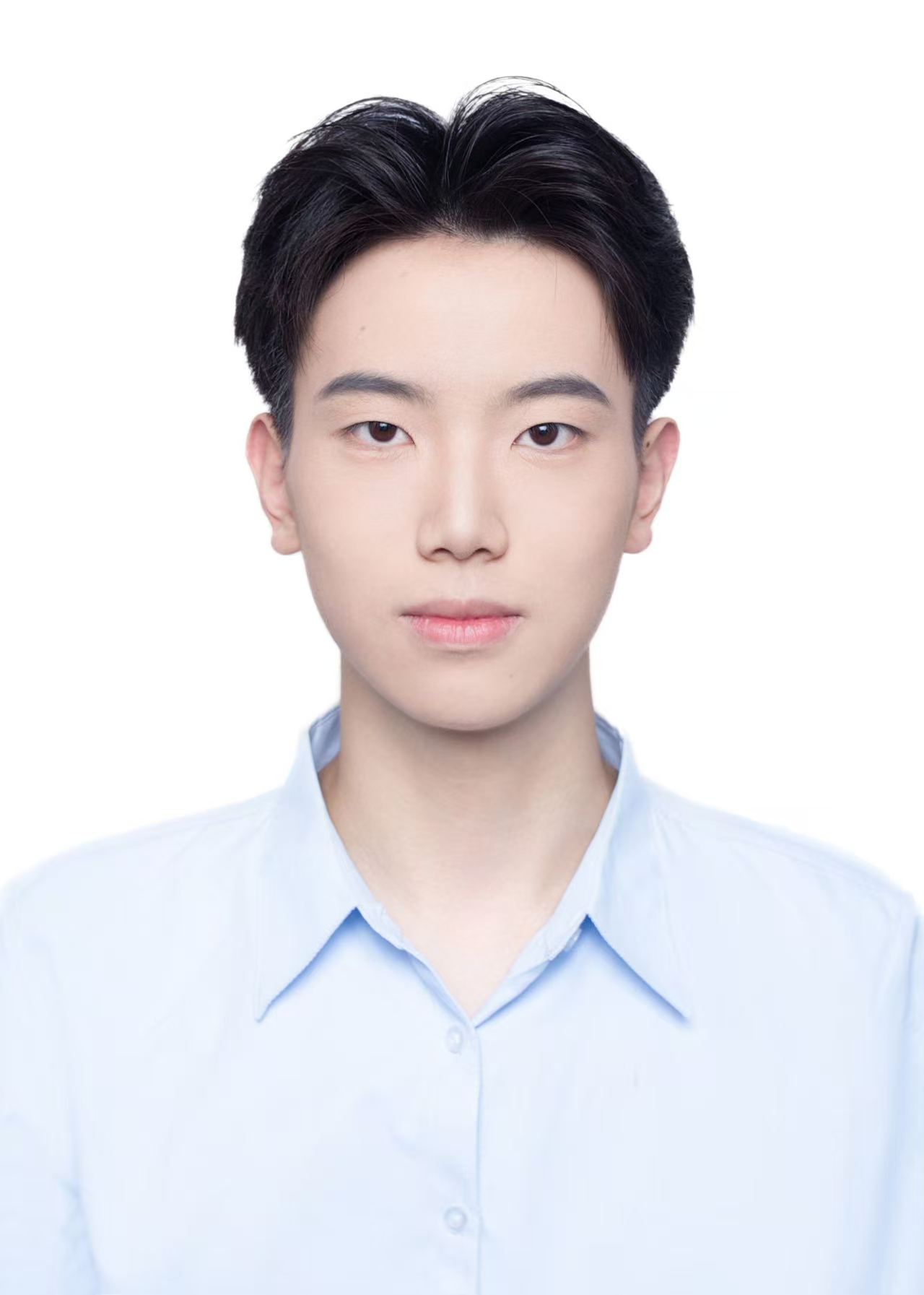}}]{Haoqian Zhang}
is currently an undergraduate student majoring in Cyberspace Security at Sichuan University, Chengdu, China. His research interest include information retrieval, recommendation system and machine learning.
\end{IEEEbiography}
\begin{IEEEbiography}[{\includegraphics[width=1in,height=1.25in,clip,keepaspectratio]{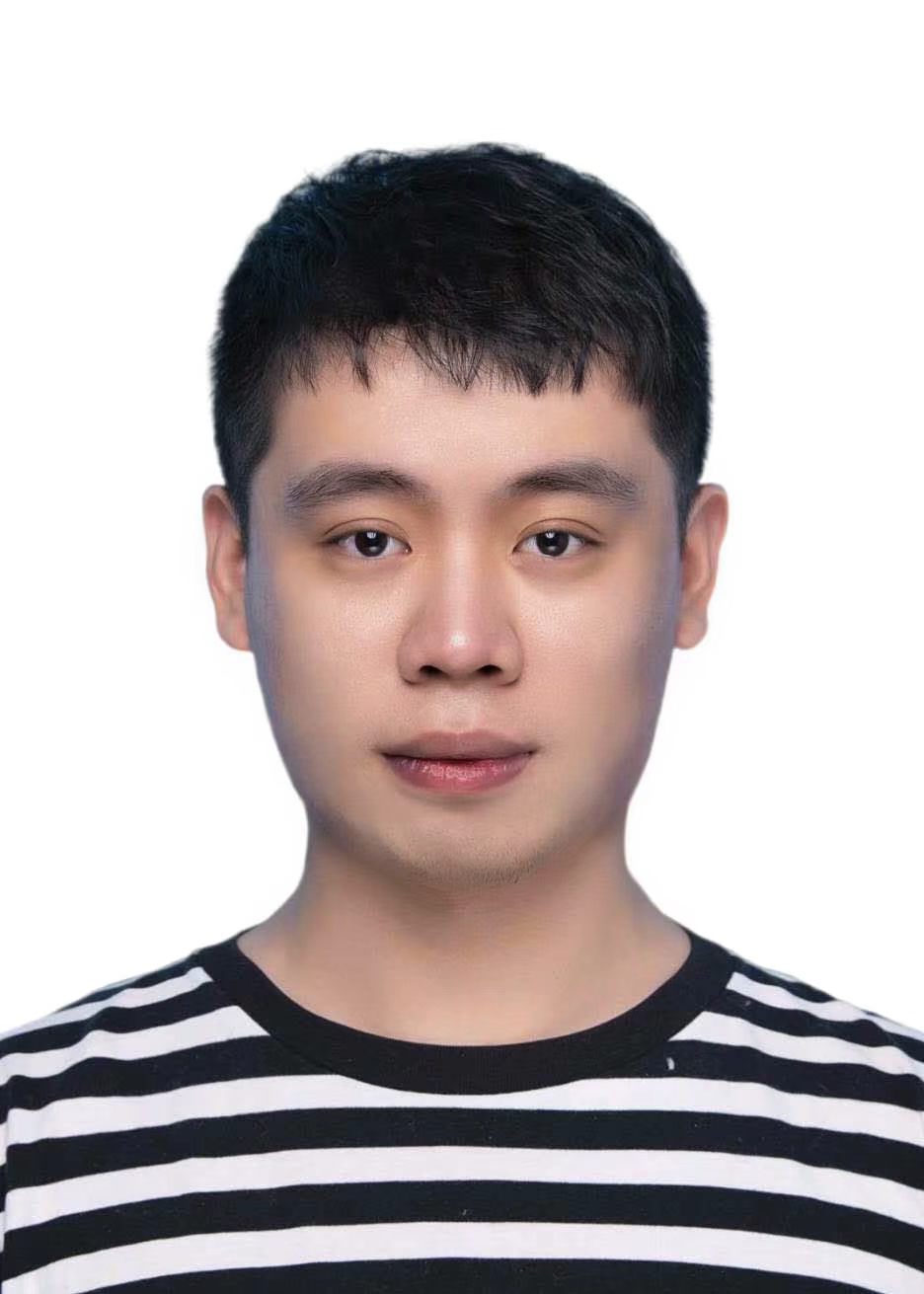}}]{Ziyuan Yang}
received the Ph.D. degree in computer science from the College of Computer Science, Sichuan University, China, in 2025.,He was a Research Intern at the Centre for Frontier AI Research, Agency for Science, Technology and Research (A*STAR), Singapore. He is currently a Research Fellow with Nanyang Technology University. In the last few years, he has published over 50 papers in leading machine learning conferences and journals, including CVPR, ICLR, AAAI, IJCV, IEEE T-IFS, IEEE T-NNLS, IEEE T-SMCS, and IEEE T-AI. He was a reviewer for leading journals or conferences, e.g. IEEE T-PAMI, IEEE T-TIP, IEEE T-IFS, IJCV, CVPR, and ICCV.
\end{IEEEbiography}
\begin{IEEEbiography}[{\includegraphics[width=1in,height=1.25in,clip,keepaspectratio]{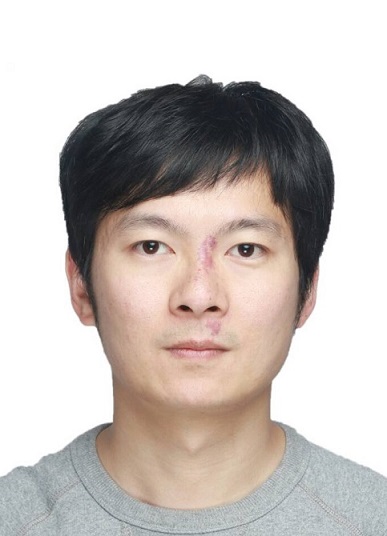}}]{Yi Zhang}
(Senior Member, IEEE) received the B.S., M.S., and Ph.D. degrees in computer science and technology from the College of Computer Science, Sichuan University, Chengdu, China, in 2005, 2008, and 2012, respectively. From 2014 to 2015, he was with the Department of Biomedical Engineering, Rensselaer Polytechnic Institute, Troy, NY, USA, as a Postdoctoral Researcher. He is currently a Full Professor with the School of Cyber Science and Engineering, Sichuan University, and is the Director of the deep imaging group (DIG). His research interests include medical imaging, compressive sensing, and deep learning. He authored more than 80 papers in the field of image processing. These papers were published in several leading journals, including IEEE TRANSACTIONS ON MEDICAL IMAGING, IEEE TRANSACTIONS ON COMPUTATIONAL IMAGING, Medical Image Analysis, European Radiology, Optics Express, etc., and reported by the Institute of Physics (IOP) and during the Lindau Nobel Laureate Meeting. He received major funding from the National Key R\&D Program of China, the National Natural Science Foundation of China, and the Science and Technology Support Project of Sichuan Province, China. He is a Guest Editor of the International Journal of Biomedical Imaging, Sensing and Imaging, and an Associate Editor of IEEE TRANSACTIONS ON MEDICAL IMAGING and IEEE TRANSACTIONS ON RADIATION AND PLASMA MEDICAL SCIENCES. 
\end{IEEEbiography}

\end{document}